# Advancing Behavior Engineering: Toward Integrated Events Modeling

Sabah Al-Fedaghi,
*sabah.alfedaghi@ku.edu.kw*
Computer Engineering Department, Kuwait University, Kuwait

**Summary**
The term behavior engineering (BE) encompasses a broad integration of behavioral and compositional requirements needed to model large-scale systems. BE forms a connection between systems-engineering processes and software-engineering processes. In software engineering, interpreting requirements can be perceived as specifying behavior, which is viewed in terms of chronology of events in the modeled system. In this paper, we adopt BE in its general and integrating sense to search for a unifying notion of an event as a fundamental behavior-modeling concept. We examine several bodies of research with various definitions of an event and its basic units and structure. We use the thinging machine (TM) model to analyze notions related to events, including Dromey's behavior trees, fluents (change over time), recurrent events, and Davidson's events. The results point to an underlying meaning that can lead to a unifying event concept.

*Key words:*
*software engineering, systems engineering, behavior, event, conceptual modeling*

## 1. Introduction

Behavior engineering (BE) [1] creates a link between systems engineering processes and software engineering processes. According to some researchers [2], BE is highly effective in practical, industry-based situations when applied to large complex systems. The present use of the term BE embraces a broader rigorous formalization and integration of large sets of behavioral and compositional requirements needed to model large-scale systems. In BE [3], a behavior model is developed systematically from requirements "in such a way that issues with consistency and completeness are revealed and resolved as the tree is built. The resulting [behavior] tree expresses all the scenarios and use cases that are implied by the requirements in a single coherent model" [4].

In software engineering, the clarity of a system's description and the absence of ambiguity are essential in specifying software requirements. Interpreting requirements can be viewed as specifying behavior [5]. The behavioral aspect of a model is viewed in terms of events as "the representation of a fact that participates in reactions of the modelized system. It occurs in a spontaneous random manner (in the case of external event) or is generated by the application" [6]. A unifying notion of events in the context of dynamic systems modeling is important for behavioral modeling. Such an achievement would advance and extend the field of behavior engineering as a link between systems-engineering processes and software-engineering processes.

In this paper, we adopt BE in its general and integrating sense to search for a unifying notion of an event as a fundamental behavior-modeling concept. We examine several bodies of research with various definitions of an event and its basic units and structure. Throughout the paper, we will apply thinging machine (TM) modeling [7-16] to analyze some concepts related to events.

1.1 A Glimpse at Some Definitions of Events

Similar to objects, events belong to categories and have properties and parts. What is an event? According to Huang and Chuang [17], an event's definitions include the following:
- Something that happens at a given place and time.
- Any physical, social, or mental process.
- A processual entity that is the fiat or bona fide instantaneous temporal boundary of a process.
- An occurrence of actions and changes in the real world.
- A perduring entity that unfolds over time.
- An action or occurrence taking place at a certain time at a specific location.
- A thing that has happened/is scheduled to happen.
- An event cannot be assigned a specific definition. Events encompass everything that happens, even fictional events.

Events are typically distinguished from objects. For example, events are exemplified by births or meetings and objects by people and organizations that (may) experience these events. As with objects, events can have attributes that describe their properties and qualities.

Some of the definitions of an event in software engineering are:

- A function that responds to an action the user or the system itself takes, e.g., the click event. A function is a group of related statements that perform some action [18].
- An action or occurrence such as a mouse click, a keystroke, mouse movements, or any system-generated notification [19].
- A significant change of state, e.g., an order in an ecommerce site or when a user views a web page [20].

1.2 Outlines of the Paper

The next section includes a brief review of our main tool to model events. The remaining sections involve examining conceptualizations of events as follows:
  Section 3: [21]'s behavior-tree model
  Section 4: The murder of Caesar as an event
  Section 5: Events and fluents (in artificial intelligence)
  Section 6: Modeling events in a geospatial domain
  Section 7: Repetition of an event (recurrent events)
  Section 8: Davidson's events

## 2. Thinging Machine Modeling

The main TM thesis is that any entity has a double nature as (i) a thing and (ii) a process (abstract machine); thus, we call these thing/machine entities *thimacs*. In TM modeling, intertwining with the world is accomplished by integrating these two entities' modes of being.

Thimacs inhibit traditional categorization, properties, and behavior, replacing them with the five actions: creating, processing, releasing, transferring, and receiving. Such a thesis implies that all actions in a system can be reduced to these five generic (elementary) actions. Since generic events are time-injected actions, there are five generic events to be explained next. Events are thimacs in conceptual space-time; conceptual space includes concepts as thimacs.

Because machines are things, all things can be reduced to five elementary things. In TM modeling, a thing's machine operates on other things by creating, processing, releasing, transferring, and/or receiving them. The term "machine" refers to a special abstract machine (see Fig. 1, simplified in Fig. 2). Among the five stages, flow (represented by the solid arrows in Fig. 1) signifies conceptual movement from one machine to another or among a machine's actions. An enduring thing is a static thimac (a thing + its machine specification) and becomes a dynamic event thimac if it embeds a time subthimac. Russell interpreted "the enduring thing or object of common sense as a world-line, a causally related sequence of events, and…it is events and not substances that we perceive." Even though Russell distinguishes between eternal objects and actual objects, he does not distinguish between an enduring thing without time and one with time. For example, thimacs without time include the Pythagorean intelligible structures and Aristotle's unmoved mover, i.e., pure form. However, Aristotle considered a sensible object to consist of both matter and form, neither of which can exist without the other. In TM, so-called objects as "matter and form" formations have two modes: static thimac (thing/machine) and dynamic (event-based) thimac. The distinction between staticity and dynamism is essential for modeling, as will be exemplified later in this paper. Thimacs without time form a system's static specification, while time is an indispensable ingredient for describing its behaviour.

The TM's actions (also called stages) can be described as follows:
- *Arrival*: A thing reaches a new machine.
- *Acceptance*: A thing is permitted to enter the machine. If arriving things are always accepted, arrival and acceptance can then be combined into the "receive" stage. For simplicity, this paper's examples assume a receiving stage exists.
- *Processing* (change): A thing undergoes a transformation that changes the thing without increasing the population size of things in the system. Processing may trigger the creation of a new thing; e.g., processing a thought may trigger the creation of a new thought.
- *Release*: A thing is marked as ready to be transferred outside the machine.

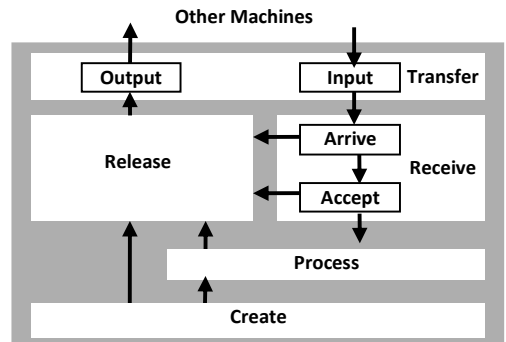

Fig. 1 The machine

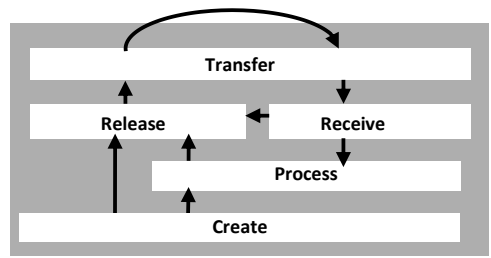

Fig. 2 The machine simplified

- *Transference*: A thing is transported somewhere outside the machine.
- *Creation*: A new thing is born (created/emerges) within a machine. Creation can be designated as "bringing into existence" in the system because what exists is what is found. Creation in a TM also means appearance in the system. Appearance here is not limited to form or solidity, but extends to any sense of the system's awareness of the new thing.

In addition, the TM model includes:
- Memory and
- Triggering (represented as dashed arrows), or relations among the processes' stages (machines); for example, the process in Fig. 1 triggers the creation of a new thing.

To approach TM modeling smoothly, we focus on the machine side of thimacs. TM modeling is a three-level process that involves the following:
- A static model of the state of affairs that produces an atemporal diagrammatic description, denoted as **S**. The state of affairs and actions are caused by the mixture of thimacs that penetrate each part (e.g., process, receive). The time of S is the present in the sense that everything subsists now.
- Decomposing S into subdiagrams that form the base of temporal events.
- The model's behavior, denoted as **B**, formulated as a chronology of events. Behavior refers to executing composite actions.

## 3. Example: Alternative Model to Behavior Tree

Dromey [21] modeled the following linguistic description:

*When a car arrives, if the gate is open, the car proceeds; if the gate is closed, when the driver presses the button, it causes the gate to open.*

Dromey [21] based his solution on the notion of a "behavior tree" as shown partially in Fig. 3. The TM model starts with the static model (Fig. 4) to arrive to the description of the behavioral model. In Fig. 4, the car arrives in the area just in front of the gate, ready to enter (circle 1). If the gate is open (2), the car proceeds (3). If the gate is closed (4), the driver presses (5) the button that causes the gate to open (2).

The behavior of this gate system is built from the static model (Fig. 4) after applying the notion of an event. An event in TM includes a time machine. For example, Fig. 5 shows the event *The car arrives in the area just before the gate, ready to enter*. The event includes the region where the event occurs in addition to other properties (e.g., intensity) that are not of concern in this paper. For simplification purposes, we will represent events by their regions. Fig. 6 shows the events in such a system as follows:

Event 1: The car arrives in the area just before the gate, ready to enter.
Event 2: The gate is open.
Event 3: The car proceeds in.
Event 4: The gate is closed.
Event 5: The driver presses the button.

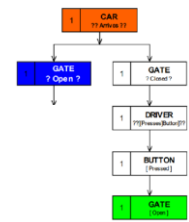

Fig. 3 Dromey's [21] behavior tree (partial)

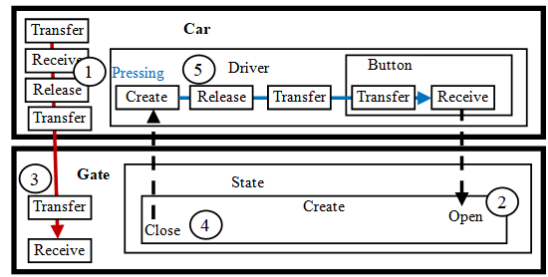

Fig. 4 The static TM model

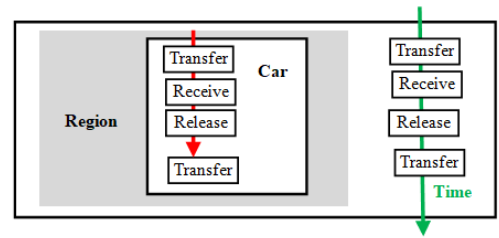

Fig. 5 The event: *The car arrives in the area to enter just before the gate, ready to enter*

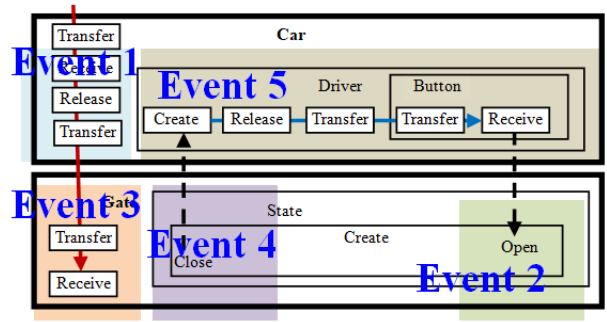

Fig. 6 Events in the TM model

Fig. 7 shows the behavioral model. Notice that repeated events in the figure (reflexive arrow) are used as a modeling convenience to denote events with the same regions (subdiagrams of the static model). Such three-stage TM modeling can be applied to all works related to the behavior tree.

## 4. The Murder of Caesar as an Event

TM modeling can be applied in a variety of systems to represent a portion of reality. According to Pegden [22], "Over the 50-year history of simulation there have been three distinct world views in use: event, process, and objects. In event worldview, the system is viewed as a series of instantaneous events that change the state of the system over time." In the TM worldview, the system is viewed as:

(1) An assembly of thimacs, including things, machines, five actions, and flows, and
(2) A chronology of events that describes the system's behavior over time.

Guizzardi and Wagner [23] give as an example the event e: the murder of Caesar. This event can be further decomposed into sub-events, namely, e1: the attack on Caesar, and e2: Caesar's death. Event e1 can, in turn, be decomposed into events e11: conspirators restraining Caesar, and e12: Brutus stabbing Caesar.

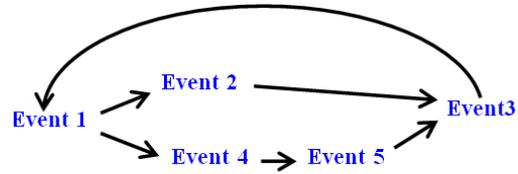

Fig. 7 The behavioral model

Events can be atomic or complex, depending on their mereological structure. While atomic events have no proper parts, complex events are aggregations of at least two disjointed events. Fig. 8 shows the TM model of the murder of Caesar. The figure shows that there is Caesar (1) who moved (2) to be in the range of the conspirators (3) who restrained him (3). In this situation, Brutus stabbed Caesar (5), triggering Caesar's death. The original creation means existence; hence, *decreation* is the reverse, in which the thing is dissolved (no longer appears in the system's view).

Such a static description includes all facts about Caesar in the given narrative. Figs. 9 and 10 show the events in the static model and the behavioral model. Note that "atomic events," in TM, are the five events that correspond to the five actions.

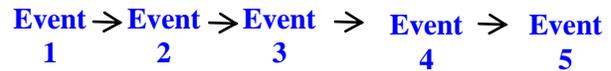

Fig. 10. The chronology of events in the murder of Caesar.

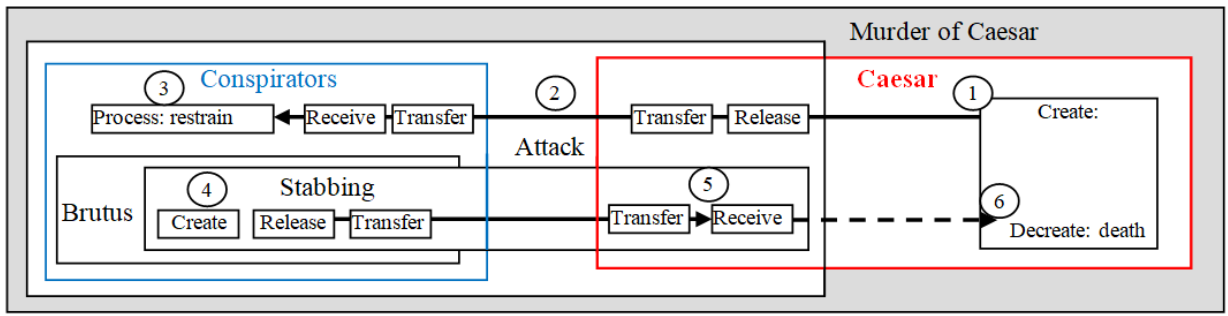

Fig. 8 The static model of the murder of Caesar

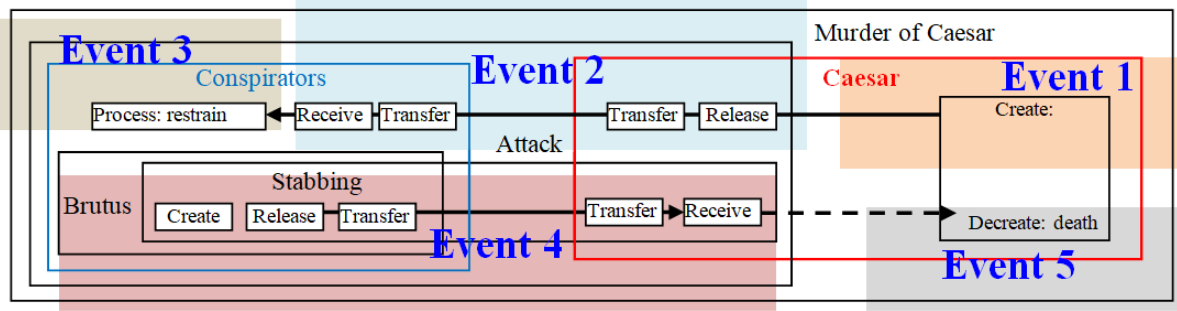

Fig. 9 The events in the model of the murder of Caesar

## 5. Fluents and Events

It seems that TM modeling can be applied to various conceptualizations of events. In artificial intelligence, a *fluent* is a condition that can change over time. For example, the condition "the box is on the table," if it can change over time, cannot be represented by the predicate On(table, box) but ON(table, box, t) where t represents time. Thielscher [24] presented fluent calculus as a specification language for robots, which met the requirements of designing robots capable of task planning on a high level. Such robots are embedded in and constantly interact with a dynamic environment. It is difficult to program suitable action sequences for all possible situations, and such an environment requires an autonomous robot to be capable of searching on its own for plans tailored to the current situation [24].

The fluent calculus is formalism for expressing dynamical domains in first-order logics in which situations are considered representations of states. For example, the action of moving the box from the table to the floor is formalized as: *State(Do(move(box,table,floor),s)) ○ on(box,table)=State(s) ○on(box,floor)* [24]. According to Thielscher [24], a fluent represents an atomic property of the physical world, which may change in the course of time. Examples of such properties can be the *location* of a movable object, the *status* of a door (i.e., whether open or closed), or the *position* of the robot.

These properties are regions in TM events. Thielscher [24] introduced a delivery scenario of a robot (see Fig. 11) using domain sorts: room, door, object, and person. Several functions are also declared, e.g.,
room |→ fluent: robot is in room;
object |→ fluent: robot carries object;
room × object × room |→ fluent: there is a request to deliver object x2 from room x1 to room x3.
Note that the fluent calculus is a second-order logic language.

This section is focused on the representation issue that includes dynamic behaviour, leaving the high-level task-planning issue for future research. Hence, we illustrate two methods to model (represent) events for the purpose of enhancing the understanding of events of dynamic systems. Fig. 12 shows the TM model that corresponds to the delivery scenario, with some modifications and limitations. The figure includes the offices (circles 1–3 in Fig. 12), storage (4), and the robot with its stations (5).

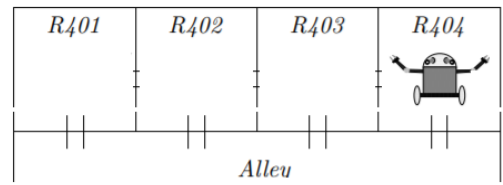

Fig. 11. A delivery scenario (partially from Thielscher [24]).

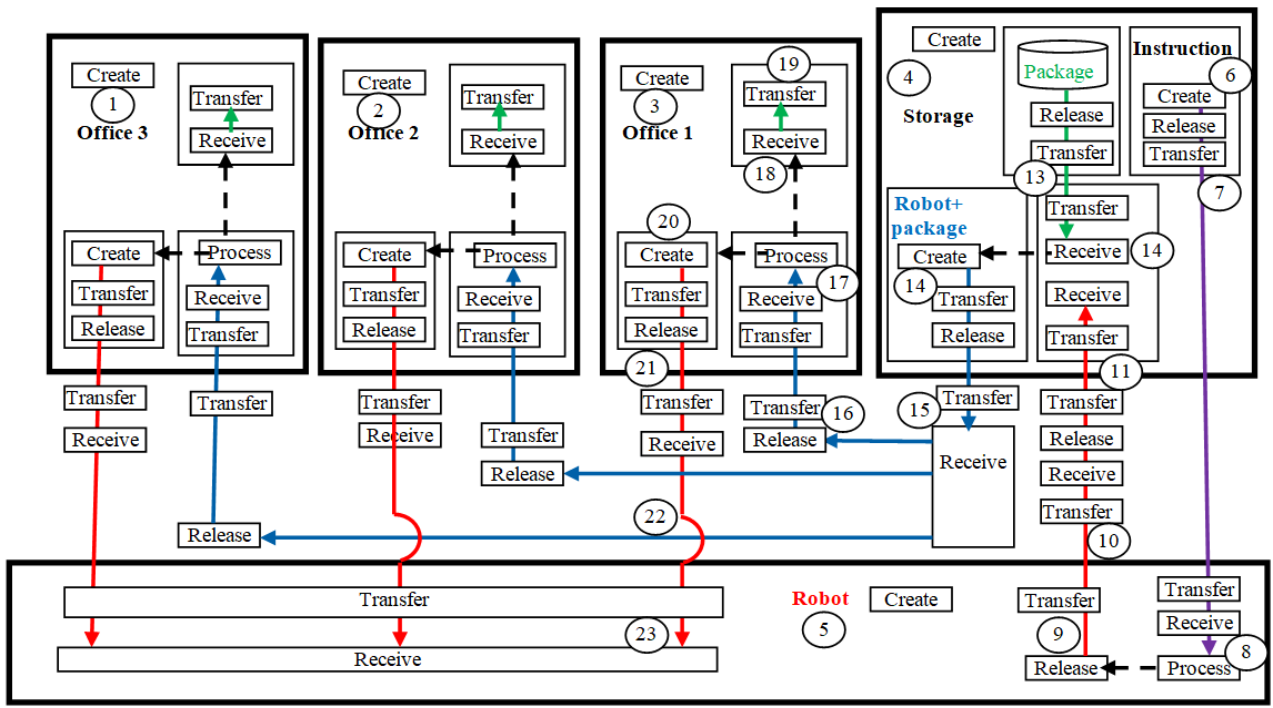

Fig. 12 The static TM model of the A delivery scenario

In the storage, instructions are sent (7) to the robot (8), where they are processed (9). Accordingly, the robot moves to the ally (9 and 10), then the storage (11). In the storage, a package is loaded (12) to the robot (13). The robot carrying the package (14) moves to the ally according to the office destination. Assuming office 1,
- The robot carrying the package moves from the ally (16) to the office (17).
- The robot delivers its package (18 and 19).
- The robot moves to the ally (20 and 21).
- From the ally, the robot moves to its station (22 and 23).

Fig. 13 shows the selected events for this delivery scenario. The events are as follows.
Event 1 ($E_1$): Delivery instructions reach the robot
Event 2 ($E_2$): The robot moves to the ally
Event 3 ($E_3$): The robot leaves the ally
Event 4 ($E_4$): The robot moves to the storage
Event 5 ($E_5$): The package is loaded in the robot
Event 6 ($E_6$): The robot carries the package
Event 7 ($E_7$): The robot moves to the ally
Event 8 ($E_8$): The robot moves to office 1
Event 9 ($E_9$): The robot delivers the package
Event 10 ($E_{10}$): The robot leaves office 1
Event 11 ($E_{11}$): The robot moves to office 2
Event 12 ($E_{12}$): The robot leaves office 2
Event 13 ($E_{13}$): The robot moves to office 3
Event 14 ($E_{14}$): The robot leaves office 3
Event 15 ($E_{15}$): The robot moves from the ally to its station in the diagram are not marked. Fig. 14 shows the behavioral model of the delivery system.

The robot performs only three types of behaviors (plans) as shown in Fig. 15. Such a description is reasonable for a department with tens of offices. However, the system should be supplemented with exceptional events. For example:
- The office door is closed: This situation can be handled within the system by returning the robot to storage to return the package.
- The office door is closed and the storage door is closed: This situation can be handled with the robot going back to its station.
- The employment of multiple robots with multiple types of things to carry. It is possible to build a traffic system using the same modeling method.

The point here is that such techniques to handle exceptions are almost sub-plans of the three established plans. The remaining questions concern the practicality of this exhaustive modeling method for handling large systems, e.g., thousands of offices. Another issue is related to complementing this model with the type of logic machine developed in Thielscher's work [24]. Each of these two modeling approaches can benefit from developments in the other, especially with regard to unifying the conception of events.

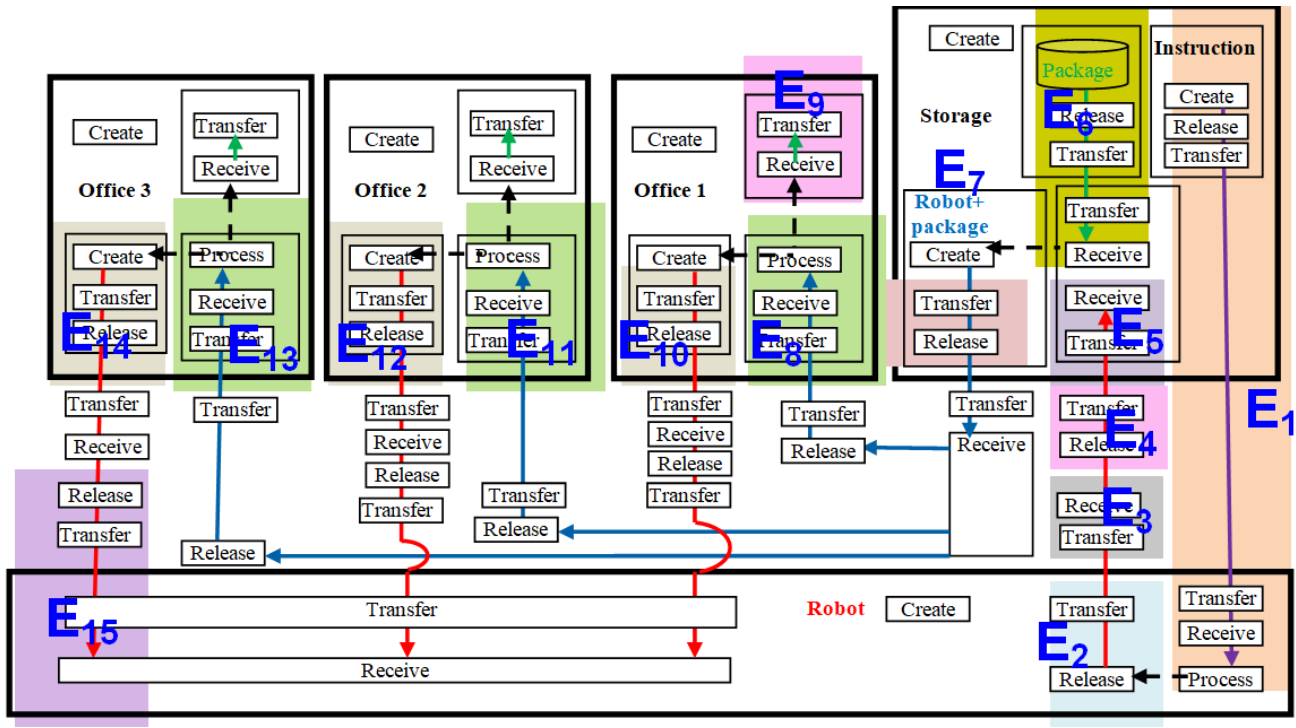

Fig. 13 Events in the delivery scenario

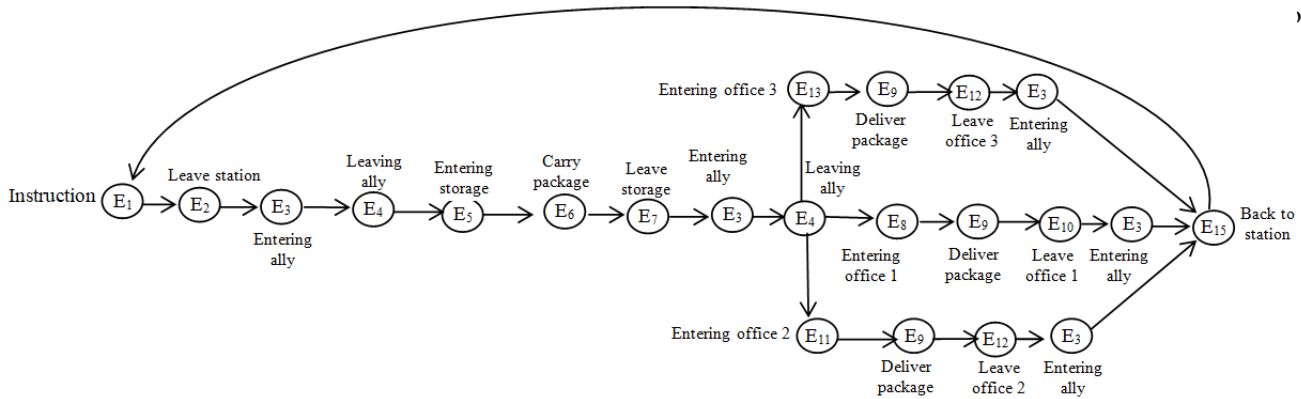

Fig. 14 The behavior of the delivery system

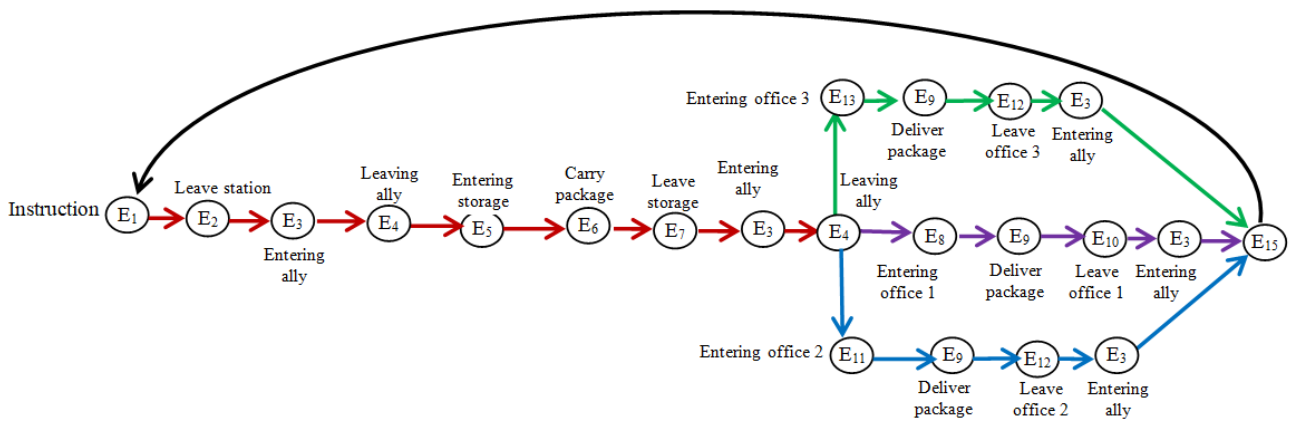

Fig. 15 The three types of plans in the delivery system where the first part is common for different plans

## 6. Modeling Events in Geospatial Domains

Hornsby and Cole [25] modelled the dynamic happenings entities experience in a geospatial domain as events; from an analysis of these events, they showed how meaningful information about objects' movement can be abstracted. A set of eight possible event pattern types are distinguished and discussed. Modeling events provides a foundation for distinguishing particular semantics of movement based on patterns of events, forming a basis for querying different kinds of events and developing automated event notification systems.

Hornsby and Cole [25] modelled a geospatial domain that is partitioned according to fiat boundaries into a set of discrete, contiguous, and non-overlapping zones. They present a harbor domain (See Fig. 16) where navigational outbound traffic lanes' separation schemes. The zones provide a reference for defining objects' spatial locations and the events associated with those objects. As an entity moves and experiences various occurrents, "describing movement from the perspective of the occurrents rather than the more typical location-based approach affords a new view of moving entities where the semantics of occurrents are the prime focus," said Hornsby and Cole [25]. Occurrents are modelled as events to generate an event-based perspective of an object's movement in a geospatial domain. An example of a representative sequence of events is:

$$E = \{{}^{v}_{ID}e^{zone}_{t1}, {}^{v}_{ID}e^{zone}_{t2}, \ldots, {}^{v}_{ID}e^{zone}_{tn}\}$$

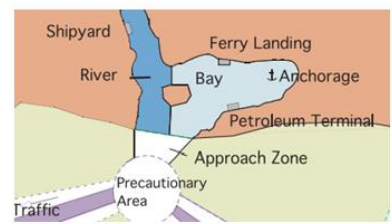

Fig. 16 Sample harbor zones (Partially from Hornsby and Cole [25])

Representing a moving object's path as a sequence of changing zone events is a rather basic semantic description of an object's movement. Such a path can portray richer depictions of object movement incorporating more semantics by including more refined types of events, referred to as specialized events. Hornsby and Cole [25] model events as being instantaneous, that is, events have no duration and mark a change in the state of an object.

To illustrate the TM modeling of changing zone events, we simplify the zones map to involve six zones as shown in Fig. 17. Additionally, to save space, we explain the model in the events diagram since it is easy to extract the static diagram by eliminating all events. In the figure, each zone has three events,
- entering the zone (transfer and receive),
- leaving the zone, and (release and transfer).
- being in the zone (processing as being in the zone takes its course).

All events in TM modeling are based on generic events (grounded on generic actions) in which events have duration. The instantaneity notion is applied only to the transition between generic events. In TM, the exact time between transfer/output (in the source machine) and transfer/input (in the destination machine) can be calculated from the end point that includes the transfer out and transfer input. The two transfers represent the boundary between zones. We can consider the object in a target zone if any part of the object is inside that zone. This is similar to running competitions in which a runner finishes the race (distance) as soon as their foot (not their whole body) touches the finishing line.

From Fig. 17, we can identify sequences of events, for example, $E_1$, $E_2$, $E_4$, $E_5$, $E_6$, $E_7$, $E_8$, $E_9$, $E_{15}$. $E_{14}$, and $E_{16}$, that describe the sequence of events for an object coming from the open sea and ending in the destination ferry landing. All types of analysis that Hornsby and Cole [25] discussed can be applied to the TM model. It seems that such a TM model is suitable for this type of application.

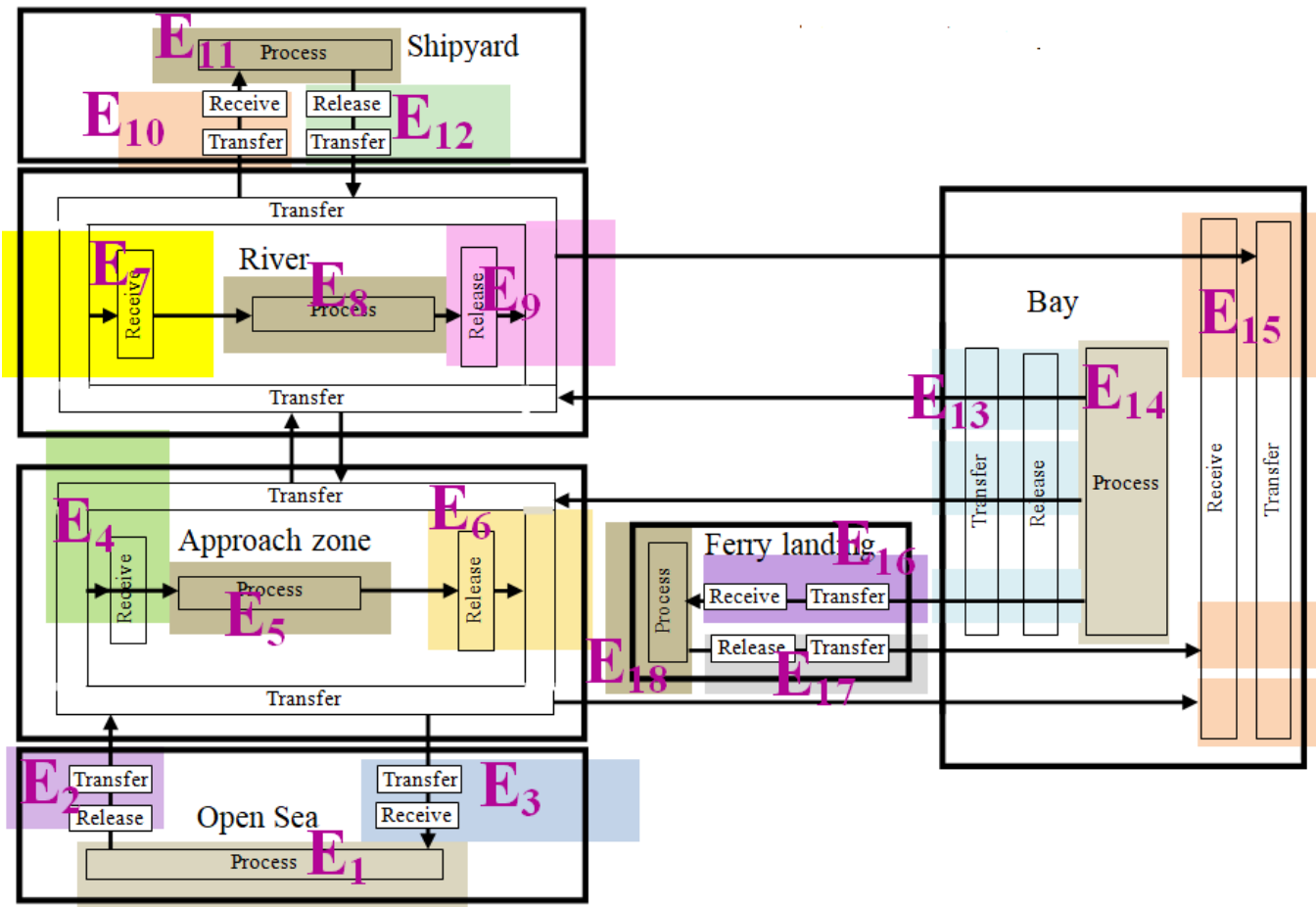

Fig. 17 The event model of six zones.

## 7. Repetition of an Event (Recurrent Events)

Carriero et al. [26] described an ontology design pattern for modeling events that recur regularly over time and share some invariant factors that unify them conceptually. Recurrent events are seen as collections of events of the same type. Such a recurrent event is represented as "both a collection of events and a situation in which these events are contextualized and unified according to one or more properties that are peculiar to each event, and occur at regular intervals" [26].

7.1 Endurants and perdurants

Carriero et al. [26] used Bottazzi et al.'s [27] concept of collection to categorize objects as:
- Endurants: e.g., physical (a hammer), nonphysical, social (an organization), or mental (a belief) objects.
- Perdurants: e.g., event (a departure) or stative (sitting).

According to Casati and Varzi [28], the variety of the world seems to lie not only in the assortment of its animals and physical objects—and perhaps minds but also in the sort of things that happen to or are performed by them. Entities of this sort appear in many situations; e.g.,
- Infants appear to be able to discriminate events.
- Humans and animals appear to plan and execute actions.
- Linguistic devices (e.g., verb tenses) are tuned to events and event structures.
- Thinking about the temporal and causal aspects of the world requires events and their descriptions. [28]

However, there is significant disagreement concerning the precise nature of such entities. Defining events as "things that happen" merely shifts the focus to clarifying the meaning of "happen." Some philosophers treat both objects and events as entities of the same kind: an object would be a "monotonous" event, and an event would be an "unstable" object [28].

In a TM, an event is a thing with a time submachine. A real event happens in (physical) time and, say, a fairy tale event happens in fairy tale time. Newtonian laws are events in which the time of their regions is the "all time" of the Newtonian world.

Further, it is important to note the notion of events implied in the used terms. Endurants are of two types, static or atemporal things (e.g., a single image [static picture]) and things that do not change over time. For example, consider a single image in a film appearing repeatedly while the film is running forward or backward with no change of the image. In the TM view, the model has many events with the same region (the image). The image is an endurant object in both situations. Analogously, world objects such as hammers, organizations, and beliefs change over time (ignoring the Whitehead's process theory) but have many events during their lifetime (hour, day, etc.). Therefore, so-called endurants have events, as in the film example mentioned previously.

Back to Carriero et al.'s [26] examples of endurants and perdurants, a *departure* is an event with a region constructed from the actions released and transferred plus the time submachine. *Sitting* is a region of an event, say, involving a person and a chair, and a time that involves transfer; received and processed (hence, there is no release and transfer of time because of *ing*). Fig. 18 shows a thing (e.g., a stone) persisting over time. Things such as stones persist through time by continuously eventizing through time. Things "endure" through time by "being wholly present [created] at more than one time" [29]. This is in contrast to the theory of persistence that claims things that last for any period of time are spatially extended as things are spread out through space—a different part for each region that the thing fills [29].

The TM modeling of these concepts is illustrated in the following two models.
1. A thing persists by repeated events that create and process (change) it—Fig. 18.
2. A thing is created and flows into time (Fig. 19). The act of creation is considered an atemporal instantaneous action that represents the *spontaneous* action of the thing's emergence. The creation is a phenomenon between a pre-"there is" and "there is." It happens in an instance.

An instant has no duration, so it is not included in the time event in Fig. 19.

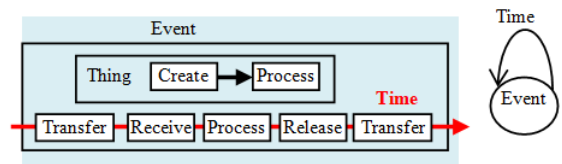

Fig. 18 The thing persists over time events.

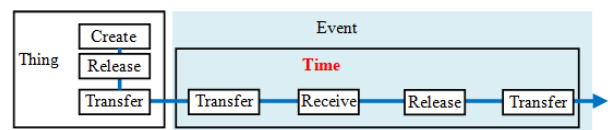

Fig. 19 The thing persists as a *change-less* thing through one event until reaching its end, e.g., is crushed.

Philosophy of religion distinguishes between creation as the action that brings a thing into existence and conservation as the action that maintains the existence of a thing over time. Conservation is continuous creation [30]. "We call the act 'creation' if it occurs at the first time at which the creature exists, and we call it 'conservation' if it occurs at a later time, but the action is the same" ([30] quoting Quinn [31]). Conservation must be an on-going act, whereas creation occurs in an instant.

In TM, the terms "conservation" and "processing" correspond to the event creation process that is continuously repeated, in which a process triggers creation. The point here is that the notions of endurants and perdurants may be analyzed in the TM modeling of these concepts, which may shed some light on these concepts as a topic for future research.

### 7.2 The event *Umbria Jazz*

Returning to Carriero et al. [26]'s ontology pattern for modeling events that recur regularly, they provided a graph consisting of a network of ontologies and facts on Italian cultural properties that represent events regularly recurring over time. *Umbria Jazz* is a recurrent event series, as shown partially in Fig. 20.

According to our understanding (and to further illustrate the TM representation), we add two subthings: workshops 1 and 2. Fig. 21 shows the static description of *Umbria Jazz*, Fig. 22 divides it into three events, and Fig. 23 shows its chronology of events. *Umbria Jazz* is one recurrent event with its unifying factors and other attributes and subthings.

## 8. Related Work (Davidson's Events)

An important role in linguistic theory is Davidson's [32] claim that events are spatiotemporal things.

### 8.1 Transitive Verbs

Maienborn et al. [33] gave a pre-Davidsonian example of a transitive verb introducing a relation. *Jones buttered the toast*, thus yielding the logical form *BUTTER (jones, the toast)*. According to Maienborn et al. [33], it seems that the sentence and its logical form reflect one event. However, such an assumption fail to consider the structure of the involved event. This would lump information in the language.

Fig. 24 shows the TM static representation of *Jones buttered the toast*.

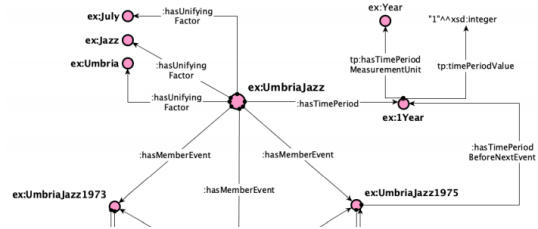

Fig. 20 Umbria Jazz is a recurrent event series, (partial, from Carriero et al. [26])

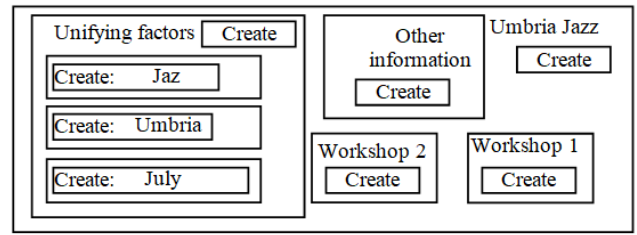

Fig. 21 An extended static model of the Umbria Jazz that includes information about the thing and subthings workshops 1 and 2.

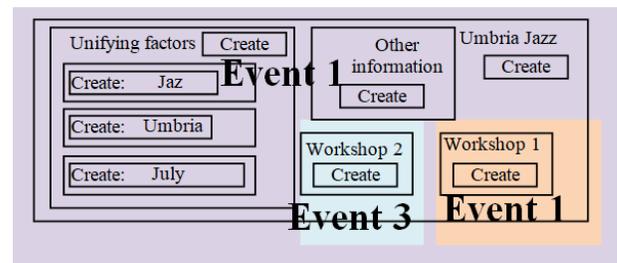

Fig. 22 Umbria Jazz is divided into three events.

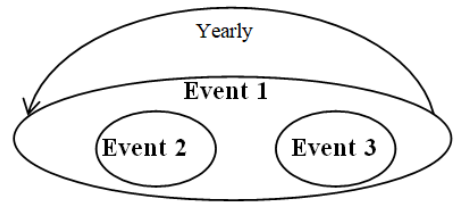

Fig. 23 Umbria Jazz as a recurrent event that includes two sub-events.

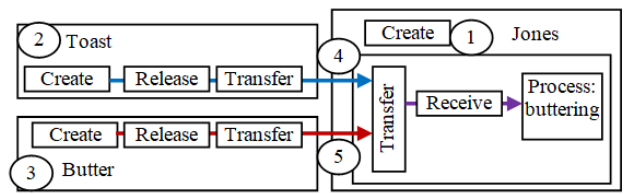

Fig. 24 The static TM representation of *Jones buttered the toast*.

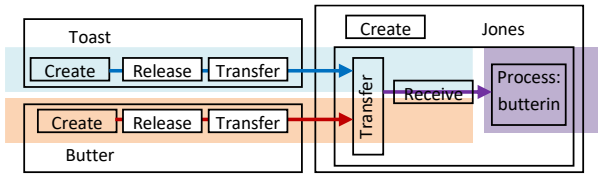

Fig. 25 The events of the TM representation of *Jones buttered the toast*.

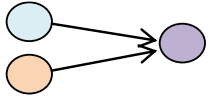

Fig. 26 The chronology of events in *Jones buttered the toast*.

First, there are Jones (1), toast (2), and butter (3). The toast and butter enter Jones's realm, expressed in TM in terms of their flow (3 and 4) to John's machine. Figs. 25 and 26 show selected meaningful events in the example and the chronology of events, respectively.

The TM diagrammatic representation seems to be richer than the logical one. However, whether the diagrammatic representation is susceptible to the reasoning process remains to be investigated. Here, we point to possible further research work. Currently, we are interested only in exploring the application of TM in Davidson's [32] ideas of the notion of events.

### 8.2 Implicit Actions

Davidson [32] pointed out that such a logical expression of the example does not allow us to refer explicitly to the action the sentence describes and specify it further by adding, e.g., that Jones did it slowly, deliberately, with a knife, in the bathroom, at midnight (Maienborn, et al. [33]). Accordingly, Davidson proposed that, *Jones buttered the toast in the bathroom with the knife at midnight* takes the logical form, $\exists e \ [BUTTER \ (jones, the \ toast, e) \ \& \ IN \ (e, the \ bathroom) \ \& \ INSTR \ (e, the \ knife) \ \& \ AT \ (e, midnight)]$ [33]. Figs. 27 and 28 show the static and events TM models of this last logical form.

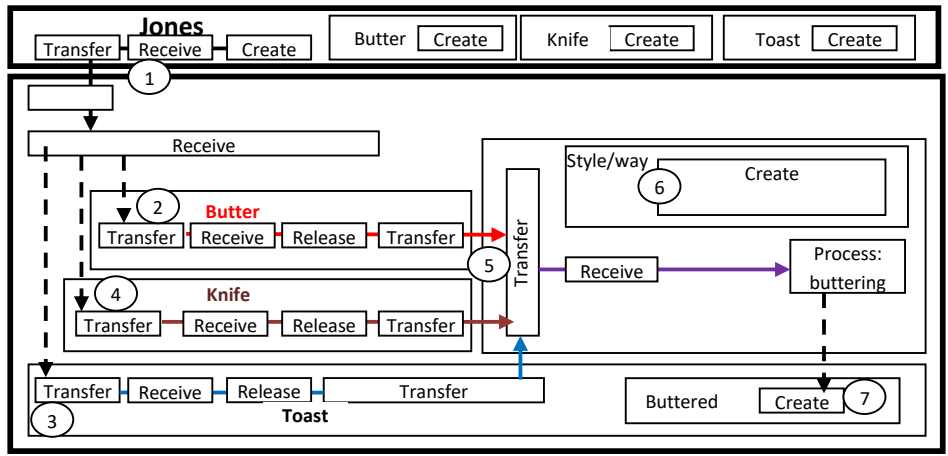

Fig. 27 The static model of $\exists e \ [BUTTER \ (jones, the \ toast, e) \ \& \ IN \ (e, the \ bathroom) \ \& \ INSTR \ (e, the \ knife) \ \& \ AT \ (e, midnight)]$

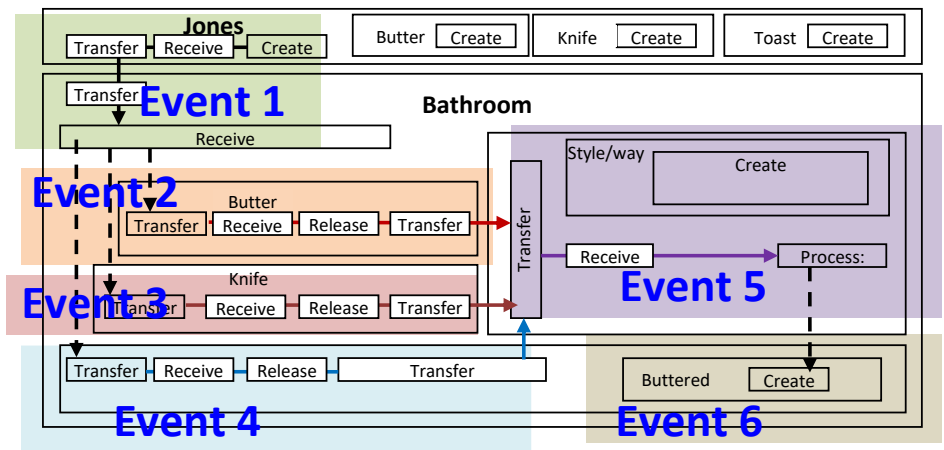

Fig. 28 The events in $\exists e \ [BUTTER \ (jones, the \ toast, e) \ \& \ IN \ (e, the \ bathroom) \ \& \ INSTR \ (e, the \ knife) \ \& \ AT \ (e, midnight)]$

In Fig. 27, Jones goe to the bathroom with the butter, toast, and a knife (1). Their arrival in the bathroom triggers (means) the arrival of the butter, toast, and knife (1, 2, and 3, respectively) in the bathroom. This begins a buttering process (5) that is slow and deliberate (6) to produce buttered toast (7).

Fig. 29 shows the chronology of these events. Event 1's time is midnight. Again, the TM diagrammatic representation seems to expose different aspects of the whole linguistic analysis of events. For example, our discussion in this field invites investigation of whether the diagrammatic representation is susceptible to the reasoning process.

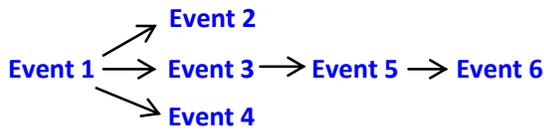

Fig. 29 The chronology of events in ∃e [BUTTER (jones, the toast, e) & IN (e, the bathroom) & INSTR (e, the knife) & AT (e, midnight)]

### 8.3 Identical Actions

Maienborn et al. [33] also discussed the issue of two events that are identical because they occupy the same portion of space and time. They take Davidson's example, in which we "wouldn't be able to distinguish the event of a metal ball rotating around its own axis during a certain time from an event of the metal ball becoming warmer during the very same time span" [33]. Figs. 30 and 31 show the TM static and events models of this case. The three events occur simultaneously; however, we can indicate the obvious causal relationship between them as shown in Fig. 32. The three events occupy different conceptual regions. They happen simultaneously with the causal relationship repeatedly.

The TM modeling seems to open a new type of event analysis such as the last example, which admittedly needs further examination.

### 8.4 Hidden Events

Maienborn et al. [33] also discussed the issue of a hidden event. A sentence such as *Anna saw Heidi cut the roses* expresses that Anna perceived the event of Heidi cutting the roses. This does not imply that Anna was necessarily aware of, e.g., who was performing the action *Anna saw Heidi cut the roses*. Fig. 33 shows the corresponding TM model. First, there are people (1), Anna (2) and Heidi (3). The image of a person (4), which could be the image of Heidi, appears to cut roses (5). Anna perceives this image (6). She then processes it (7) to recognize Heidi (8).

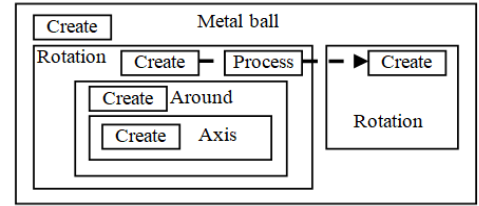

Fig. 30 TM model of a metal ball rotating around its own axis and becomes warmer

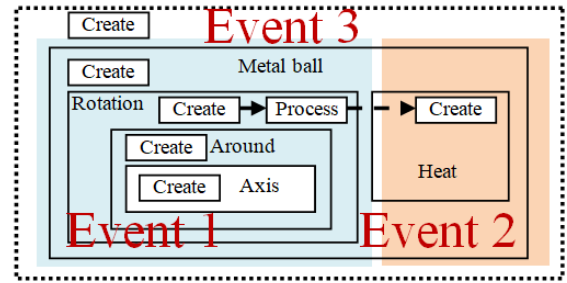

Fig. 31 TM events of a metal ball rotating around its own axis and becomes warmer

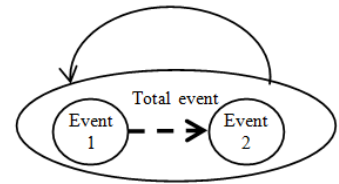

Fig. 32 The causal relationship between the two events

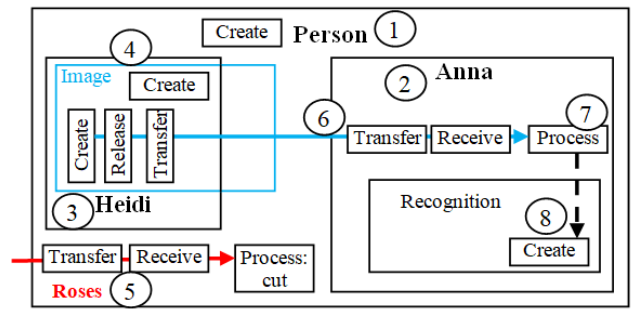

Fig. 33 TM model of Anna saw Heidi cut the roses

Fig. 34 shows selected events in this example.
Event 1: Appearance of a person.
Event 2: Appearance of Anna.
Event 3: Appearance of Heidi.
Event 4: A person cutting roses.
Event 5: Anna perceives the image of a person cutting roses.

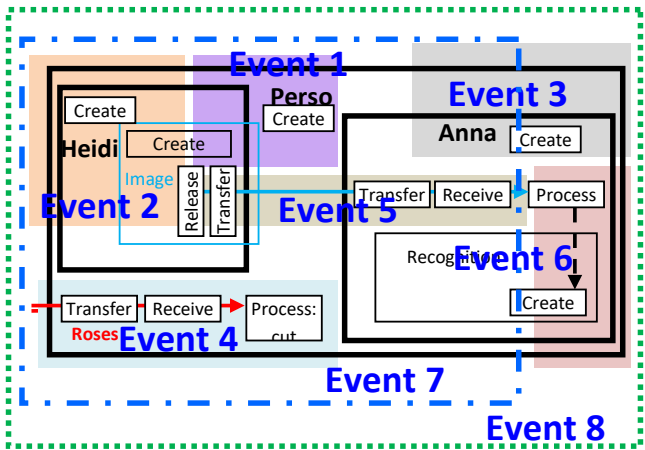

Fig. 34 Events of Anna saw Heidi cut the roses

The result points to the viability of TM as a unifying base for modeling events.

Event 6: Anna recognizes that Heidi is cutting the roses
Event 7: The union of events 1–5
Event 8: The union of events 1–6

Fig. 35 shows the chronology of events, including the two events of recognizing and not recognizing Heidi. The TM model presents a tool to represent and analyze the events, at least in the software engineering field.

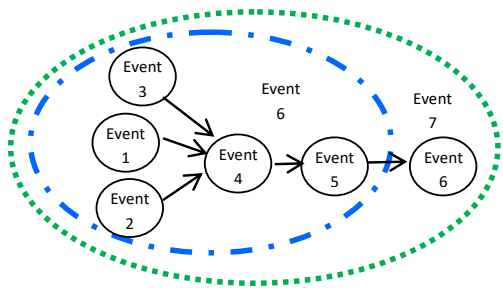

Fig. 35 Events chronology of Anna saw Heidi cut the roses

## 9. Conclusion

In this paper, we have examined several bodies of research with different definitions of events and their basic units and structure. We used the thinging machine (TM) model to analyze various notions related to events, including Dromey's behavior trees, fluents, recurrent events, and Davidson's events. The various conceptualizations of the event notion were recast using the TM model. The results point to TM's ability to express these conceptualizations. Additionally, TM has brought up some new aspects of these conceptualizations that deserve future exploration.